\documentclass[aps,prb,reprint,amsmath,amssymb,longbibliography,floatfix]{revtex4-2}
\usepackage[T1]{fontenc}
\usepackage{graphicx}
\usepackage{booktabs}
\usepackage{bm}
\usepackage[colorlinks=true,citecolor=blue,linkcolor=blue,urlcolor=blue]{hyperref}
\newcommand{\paperfigure}[2]{\includegraphics[width=#1]{figures/#2.pdf}}
\newcommand{\Sb}{S_{\mathrm b}}
\newcommand{\Tb}{T_{\mathrm b}}
\newcommand{\Eb}{E_{\mathrm b}}
\newcommand{\Es}{E_{\mathrm s}}
\newcommand{\Cb}{C_{\mathrm b}}
\newcommand{\dd}{\mathrm d}
\newcommand{\av}[1]{\left\langle #1\right\rangle}
\begin{document}
\title{Adiabatic pulsed-field dynamics of a frustrated spin-lattice system}
\author{Hidemaro Suwa}
\email{suwamaro@phys.s.u-tokyo.ac.jp}
\affiliation{Department of Physics, The University of Tokyo, Tokyo 113-0033, Japan}
\date{October 31, 2026}
\begin{abstract}
Magnetocaloric measurements in pulsed magnetic fields probe competing phases through the coupled evolution of magnetization and temperature. We formulate an adiabatic classical spin-lattice simulation that treats bond and site optical coordinates explicitly while representing the remaining thermal degrees of freedom by a finite internal reservoir. Magnetic work and energy exchange determine its temperature along each trajectory, providing feedback to the Langevin dynamics without external heat flow. Applied to a pyrochlore exchange-striction model motivated by CdCr$_2$O$_4$, the method produces a half-magnetization plateau, cooling near its onset, a temperature dome, and a change in cooling slope near plateau exit. Their joint emergence without an imposed temperature profile captures the experimental sequence of magnetic and thermal features. An approximate magnetization crossing near the temperature maximum is consistent with the reversible thermodynamic interpretation of the dome. The phenomenological reservoir enables this qualitative comparison within a simple model.
\end{abstract}
\maketitle

\section{Introduction}
The magnetocaloric effect (MCE) probes field-driven entropy redistribution through temperature changes and anomalies. Frustrated magnets are especially promising because a field reorganizes a large manifold of accessible states~\cite{Zhitomirsky2003}. The magnetic Gr\"uneisen response likewise probes entropy accumulation near quantum criticality~\cite{Zhu2003,Garst2005}. Recent advances in pulsed-field thermometry have expanded access to this response, with thin-film measurements up to 55~T~\cite{Kihara2013}, capacitive measurements up to 61~T~\cite{Miyake2020}, and proof-of-concept radio-frequency detection at 120~T on microsecond scales, where calibration and response delays remain important~\cite{Ogawa2026}. Simultaneous specific-heat and thermal-conductivity measurements constrain heat storage and transport~\cite{Nomoto2023}, while combined MCE and magnetostriction measurements connect thermal anomalies to magnetic and structural changes~\cite{Tang2024}. These developments make the temperature response increasingly accessible and call for models that predict its evolution together with magnetization.

Equilibrium MCE calculations reconstruct $S(T,h)$ and solve $S(T_f,h_f)=S(T_i,h_i)$. Classical Monte Carlo (MC) simulations established frustration-enhanced cooling~\cite{Zhitomirsky2003}, adiabatic demagnetization in Gd$_2$Ti$_2$O$_7$~\cite{Sosin2005}, and entropy and temperature changes in Gd$_5$(Si$_x$Ge$_{1-x}$)$_4$~\cite{Nobrega2005}. Quantum calculations resolved isentropes near gap closing and saturation~\cite{Honecker2006,Honecker2009}, and tensor networks predicted cooling from fractional excitations in Kitaev magnets~\cite{Li2024MCE}. First-principles-informed Monte Carlo simulations also treat magnetic and lattice contributions in magnetostructural alloys~\cite{Buchelnikov2011}. These are successful routes to reversible MCE when the total entropy is specified.

Finite-duration field pulses can also access nonequilibrium plateau states. In the $S=1/2$ kagom\'e magnet Cu-titmb, a transient one-third magnetization plateau appears at intermediate field-sweep rates but is absent in steady fields~\cite{Narumi2004}. In the $S=1$ organic spin ladder BIP-TENO, a one-third plateau observed with microsecond pulses is absent in slower millisecond measurements; its emergence was interpreted in terms of metastability associated with suppressed lattice relaxation during the rapid pulse~\cite{Nomura2022}. These observations show that the pulse protocol can select magnetic states whose formation and persistence depend on the competition between driving and internal relaxation. Equilibrium calculations alone cannot determine this dynamical selection.

Describing such behavior requires following the magnetization dynamics under the time-dependent field together with the lattice response and energy redistribution. Nucleation, delayed equilibration, and irreversible heating can all modify the trajectory, while a fixed-temperature thermostat omits the thermal feedback of an isolated sample. This motivates adiabatic simulation methods that evolve spins and relevant lattice coordinates while determining the internal temperature from magnetic work and energy exchange. Such a description can address both the magnetic states reached during a pulse and their accompanying magnetocaloric response.

Frustrated magnets provide a natural setting for this problem. The extensive degeneracy of the pyrochlore antiferromagnet~\cite{Moessner1998} makes its order sensitive to exchange striction, which stabilizes bond order~\cite{Tchernyshyov2002}, a half-magnetization plateau~\cite{Penc2004,Bergman2006}, and ordered distortions~\cite{Aoyama2016}, and can produce anomalous thermal expansion~\cite{Rossi2019}. Magnetoelastic measurements and modeling of MnSc$_2X_4$ ($X={\rm S,Se}$) further illustrate the importance of spin-lattice coupling~\cite{Grumbach2025}. A combined bond-site model accounts for equilibrium properties of CdCr$_2$O$_4$, although its measured temperature trajectories were not calculated~\cite{Gen2026}. The underlying ingredients, distance-dependent exchange and elastic restoring forces, are common to many magnetic materials.

The tools for such a dynamical description build on stochastic spin theory~\cite{Brown1963,GarciaPalacios1998} and atomistic integration methods~\cite{Mentink2010,Evans2014}. Explicit spin-lattice dynamics~\cite{Ma2008} connects magnetic motion to structural degrees of freedom, with distinct mechanisms for energy and angular-momentum relaxation~\cite{Strungaru2021}; spin dynamics also provides access to frustrated-magnet spectroscopy~\cite{Szabo2024}. Extensions to electronic collective variables~\cite{Chern2018,Yamasaki2026} and coherent-state geometric and Langevin dynamics~\cite{Zhang2021,Dahlbom2022Geometric,Dahlbom2022Langevin} broaden the range of accessible systems. The microcanonical Monte Carlo method offers a related approach to auxiliary energy bookkeeping~\cite{Creutz1983}, although its updates do not represent physical pulse-time evolution.

Direct simulations of thermal feedback provide particularly close precedents. Spin-lattice-electron dynamics couples explicit spins and atomic motion to an internal electronic temperature governed by heat transfer~\cite{Ma2012}. Langevin spin dynamics with first-principles parameters has simulated transient-field MCE cycles in Fe and Gd~\cite{Ma2014}, and spin-lattice calculations have quantified phonon entropy changes in Gd and MnAs~\cite{Patra2023,Patra2026}. A recent micromagnetic formulation couples stochastic magnetization dynamics bidirectionally to a spatially resolved thermal bath~\cite{Yi2026}. These studies establish dynamical MCE and evolving internal temperatures as useful approaches to magnetic thermal response.

Here we develop an adiabatic simulation framework for frustrated spin-lattice systems, separating the optical distortions that select order from the thermal background of unresolved lattice modes. Bond and site coordinates evolve explicitly, while a finite internal reservoir exchanges energy with the resolved system. Its temperature is updated from magnetic work and the resolved energy change for each trajectory, closing the feedback in the coupled Langevin dynamics.

Applied to a simple pyrochlore model motivated by CdCr$_2$O$_4$, this framework captures the experimental sequence of a half-magnetization plateau, cooling near its onset, a temperature dome, and a change in cooling slope near plateau exit~\cite{Gen2026}. The joint appearance of these magnetic and thermal signatures, without prescribing the temperature profile, is the central qualitative result. The near crossing of magnetization curves around the temperature maximum also connects the dynamics to an equilibrium interpretation of the dome. Our focus is this qualitative correspondence within a classical model and a phenomenological reservoir; material-specific heat capacities and temperature amplitudes remain targets for future extensions.

The paper is organized as follows. Section~\ref{sec:method} presents the model, coupled dynamics, and thermodynamic framework. Section~\ref{sec:simulation} describes the simulation parameters and experimental data. Section~\ref{sec:results} compares the magnetic and thermal responses with experiment and a low-temperature isothermal reference. Section~\ref{sec:implications} discusses relaxation, nucleation, and implications for experiments and other materials, followed by the conclusions in Sec.~\ref{sec:conclusion}. Appendixes~\ref{sec:frequency} and \ref{sec:validation} examine optical-frequency dependence and numerical consistency, respectively.

\section{Adiabatic spin-lattice dynamics}\label{sec:method}
\subsection{Resolved spins and optical coordinates}
We consider unit classical spins $\bm S_i$ on a pyrochlore lattice, with field direction $\bm n\parallel[111]$. The field variable $h$ has units of energy and couples to $M=\sum_i\bm n\cdot\bm S_i$. The normalized magnetization is $m=M/N$, with saturation value unity. Optical distortions modulate the nearest-neighbor exchange as
\begin{equation}
 J_{ij}=J-g_{\rm B}q_{ij}
 -g_{\rm S}\bm e_{ij}\cdot(\bm u_i-\bm u_j).
 \label{eq:exchange}
\end{equation}
Setting $k_{\rm B}=1$, the resolved Hamiltonian is
\begin{align}
\mathcal H_{\mathrm s}={}&
 \sum_{\langle ij\rangle}J_{ij}\bm S_i\cdot\bm S_j
 +J_{3a}\sum_{\langle ij\rangle_{3a}}\bm S_i\cdot\bm S_j \nonumber\\
 &+\sum_{\langle ij\rangle}\bm D_{ij}\cdot(\bm S_i\times\bm S_j)-hM \nonumber\\
 &+\sum_{\langle ij\rangle}\left(\frac{K_{\rm B}}2q_{ij}^2+
                   \frac{p_{ij}^2}{2m_{\rm B}}\right)
 +\sum_i\left(\frac{K_{\rm S}}2|\bm u_i|^2+
                   \frac{|\bm p_i|^2}{2m_{\rm S}}\right).
 \label{eq:ham}
\end{align}
Here $q_{ij}$ is an independent bond coordinate, $\bm u_i$ a three-component site coordinate, and $\bm e_{ij}$ the unit vector from site $j$ to site $i$. These effective optical coordinates describe local bond and site distortions.

The undistorted pyrochlore lattice has two symmetry-inequivalent third-neighbor families, both connecting sites on the same fcc sublattice at the same distance. We define $\langle ij\rangle_{3a}$ as the endpoints of two collinear nearest-neighbor bonds $i$--$k$ and $k$--$j$, with a common nearest-neighbor site at $\bm r_k=(\bm r_i+\bm r_j)/2$. The other family, $J_{3b}$, connects opposite vertices of an elementary hexagonal loop and has no intervening site at the midpoint. Each family has six neighbors per site~\cite{Jaubert2025}. We retain only the former coupling, $J_{3a}/J=-0.03$, and set $J_{3b}=0$.

The Dzyaloshinskii--Moriya (DM) vector is $\bm D_{ij}=D_0\bm d_{ij}$, with
\begin{equation}
 \bm d_{ij}=-\frac{(\bm r_i-\bm r_c)\times(\bm r_j-\bm r_c)}
 {|(\bm r_i-\bm r_c)\times(\bm r_j-\bm r_c)|},
 \label{eq:dm}
\end{equation}
where $\bm r_c$ is the center of the tetrahedron containing the bond. This convention fixes the sign of $D_0$; the interaction breaks global spin-rotation symmetry~\cite{Elhajal2005}.

The choice $|D_0|/J=0.10$ is motivated by the reported estimates $D=0.14$~meV and $J=1.35$~meV, which give $D/J\simeq0.104$~\cite{Choi2013}. Together with lattice distortion, DM twists collinear order into a spiral consistent with the measured wave vector $(0,\delta,1)$, $\delta\simeq0.09$, in reciprocal lattice units~\cite{Chung2005,Choi2013}. It therefore influences low-field order as well as relaxation.

The experimental modulation period, approximately $1/\delta\simeq11$ cubic unit cells, cannot be accommodated in our $L=4$ periodic cell. The simulated low-field reference state is commensurate and collinear. Our focus is the subsequent magnetization process during a field pulse. For the underlying bond-site model at the same $b=0.2$ and $\eta=0.6$, equilibrium magnetization curves at $T/J=0.01$ are essentially unchanged between $L=4$ and 6, including the plateau width and high-field phase~\cite{Gen2026}. This weak size dependence motivates the present cell size for a qualitative study of magnetization.

Completing the square in each optical coordinate at fixed spins gives
\begin{align}
 \mathcal H_{\rm sl}^{\rm eff}={}&
 -\frac{g_{\rm B}^2}{2K_{\rm B}}\sum_{\langle ij\rangle}(\bm S_i\cdot\bm S_j)^2
 \nonumber\\
 &-\frac{g_{\rm S}^2}{2K_{\rm S}}\sum_i
 \left|\sum_{j\in\mathrm{NN}(i)}\bm e_{ij}(\bm S_i\cdot\bm S_j)\right|^2.
 \label{eq:effective}
\end{align}
The bond term favors local collinearity; the site term couples neighboring bond correlations and selects their spatial arrangement. We use the bond-site parametrization~\cite{Gen2026}
\begin{equation}
 b=\frac{g_{\rm B}^2}{2K_{\rm B}J}+\frac{g_{\rm S}^2}{K_{\rm S}J},\qquad
 b'=\frac{g_{\rm S}^2}{K_{\rm S}J},\qquad \eta=b'/b.
\end{equation}

The negative $J_{3a}$ in Eq.~\eqref{eq:ham} is a bare exchange and must be distinguished from the interaction after lattice relaxation. For a straight nearest-neighbor path $i$--$k$--$j$, the site term in Eq.~\eqref{eq:effective} contributes $+b'J(\bm S_i\cdot\bm S_k)(\bm S_k\cdot\bm S_j)$ because $\bm e_{ki}\cdot\bm e_{kj}=-1$. For collinear unit spins, this product reduces to $\bm S_i\cdot\bm S_j$, yielding an antiferromagnetic third-neighbor contribution~\cite{Gen2026}. The net coupling in this collinear sector is therefore
\begin{equation}
 J_{3a}^{\rm eff}=J_{3a}+b'J=(-0.03+0.12)J=0.09J>0.
 \label{eq:j3-effective}
\end{equation}
Thus the bare ferromagnetic exchange used here is compatible, at the level of the net interaction's sign, with the antiferromagnetic third-neighbor exchange inferred for CdCr$_2$O$_4$~\cite{Choi2013}. Away from collinearity, the phonon-mediated term remains a three-site interaction, so this comparison does not imply an exact identification with a bilinear exchange parameter.

This effective spin interaction captures the equilibrium effect of classical harmonic distortions. To track energy storage and structural relaxation during a pulse, we retain the coordinates and their conjugate momenta explicitly. They are effective classical variables for spin-coupled distortions, rather than a mode-by-mode representation of the material's quantum phonon spectrum.

The positive spring constants $K_{\rm B}$ and $K_{\rm S}$ can be absorbed into the length scales of their respective coordinates. For a bond coordinate or a site component $x_\lambda$, with $\lambda={\rm B,S}$, define
\begin{equation}
 X_\lambda=\sqrt{\frac{K_\lambda}{J}}x_\lambda,
 \qquad \bar g_\lambda=\frac{g_\lambda}{\sqrt{K_\lambda J}}.
 \label{eq:optical-normalization}
\end{equation}
Writing its spin coupling as $-g_\lambda x_\lambda O_\lambda$, where $O_\lambda$ is the corresponding spin-dependent scalar, the oscillator contribution becomes
\begin{equation}
 \frac{\mathcal H_\lambda}{J}
 =\frac{\dot X_\lambda^2}{2\omega_\lambda^2}
  +\frac{X_\lambda^2}{2}-\bar g_\lambda X_\lambda O_\lambda.
 \label{eq:normalized-oscillator}
\end{equation}
Here $\omega_\lambda^2=K_\lambda/m_\lambda$. We report numerical frequencies as $\omega_\lambda/J$ in the spin-dynamical convention, with the precession coefficient absorbed into the time unit; the conversion to laboratory angular frequencies is given in Sec.~\ref{sec:optical-scale}. Thus the optical frequencies set the oscillation scales, while $\bar g_{\rm B}^2=2(b-b')$ and $\bar g_{\rm S}^2=b'$ set the static spin-lattice coupling. At fixed $b,b'$ and $\omega_\lambda$, different positive spring constants describe the same model after rescaling $g_\lambda$ and $m_\lambda$ along with the coordinates. The stochastic dynamics additionally requires the corresponding friction and noise rescaling, described below. We therefore specify the lattice parameters through these invariant combinations. A physical displacement scale would require a separate calibration of the effective coordinates.

\begin{figure*}
 \paperfigure{0.96\textwidth}{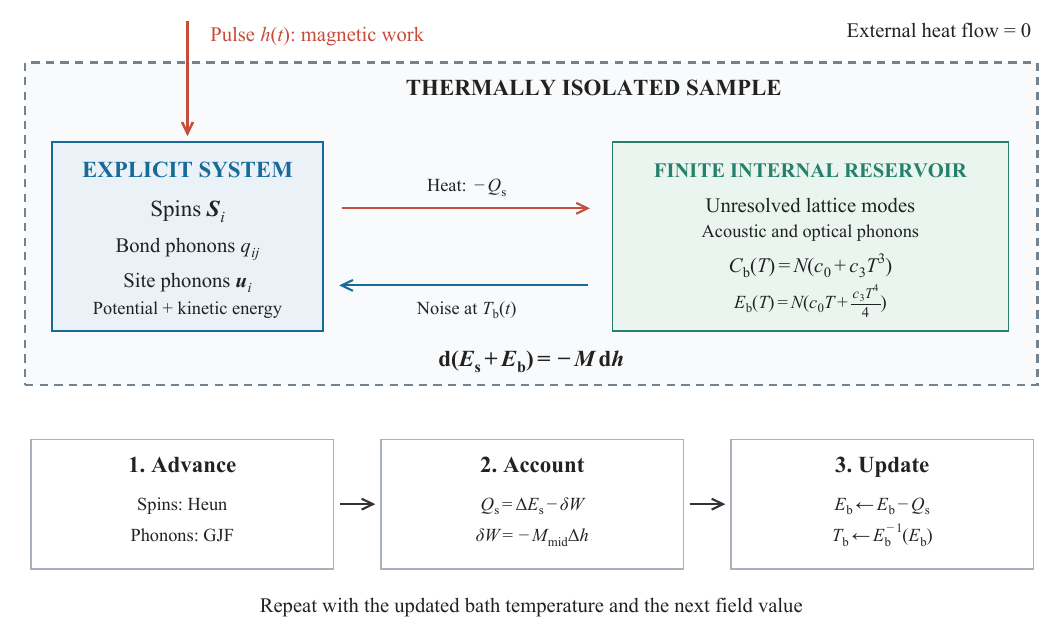}
 \caption{Adiabatic simulation of a thermally isolated spin-lattice system. Spins, bond coordinates, and site coordinates evolve explicitly and exchange heat with a phenomenological internal reservoir representing unresolved lattice modes, which can include both acoustic and optical phonons. The displayed reservoir equation of state is the phenomenological form used here; material-specific applications can incorporate the quantum heat capacity of the unresolved modes. Its temperature $T_{\rm b}(t)$ controls the Langevin noise. Magnetic work and the resolved energy change update the reservoir energy, from which its equation of state determines the next temperature. GJF denotes the optical Langevin integrator in Eqs.~\eqref{eq:gjfx} and \eqref{eq:gjfv}.}
 \label{fig:method}
\end{figure*}

\subsection{Finite internal reservoir and magnetic work}
To describe energy exchange with unresolved lattice modes, we introduce a spatially uniform phenomenological reservoir with
\begin{align}
 \Cb(T)&=N(c_0+c_3T^3),\label{eq:capacity}\\
 \Eb(T)&=N\left(c_0T+\frac{c_3}{4}T^4\right).\label{eq:eb}
\end{align}
The cubic form is motivated by low-temperature acoustic heat capacity, while its coefficient is an independent phenomenological parameter. The reservoir represents degrees of freedom distinct from the explicit optical coordinates. We assume that its modes equilibrate among themselves rapidly enough to admit a common temperature and short-memory noise, whereas the resolved spins and optical modes may relax more slowly. With $T$ measured in energy units, $c_0$ is dimensionless and $c_3$ has units $J^{-3}$; thus $C_{\rm b}/(Nk_{\rm B})=c_0+(c_3J^3)(T/J)^3$.

The reservoir can also include optical modes with negligible direct spin coupling, provided that energy exchange with other lattice modes establishes a common reservoir temperature on the relevant time scale. In the classical limit, $N_{\rm opt,b}$ unresolved harmonic modes contribute $N_{\rm opt,b}k_{\rm B}$ to the heat capacity, providing a natural interpretation of $c_0=N_{\rm opt,b}/N>0$ even in an insulator. These modes must be distinct from the explicitly resolved coordinates to avoid double counting. Here we set $c_0=0$ to simplify the reservoir to a single cubic contribution for the qualitative study; this choice neither follows from electrical insulation nor implies the absence of other optical modes. For a material-specific description, their quantum heat capacity must be considered: a constant $c_0$ is appropriate only where these modes are classically populated or their contribution is approximately temperature independent over the range of interest. Lattice heat-capacity measurements can guide this refinement. The constant $c_0$ is distinct from the electronic contribution of a normal metal, which is linear in $T$ at low temperature.

Let $\Es=\mathcal H_{\rm s}$ include optical kinetic energies and the Zeeman term. The magnetic work is $\delta W=-M\,\dd h$, so thermal isolation requires
\begin{equation}
 \dd(\Es+\Eb)=\delta W,\qquad
 \delta Q_{\rm s}=\dd\Es-\delta W=-\dd\Eb.
 \label{eq:firstlaw}
\end{equation}
Thus changing the field changes the total energy even though external heat flow vanishes. At finite sweep rates, this thermal isolation still allows irreversible entropy production. With the alternative energy $E_0=\Es+hM$, the same balance reads $\dd(E_0+\Eb)=h\,\dd M$; the energy and work conventions must be used consistently.

For each realization, the numerical update is
\begin{align}
 \delta W_n&=-\tfrac12(M_n+M_{n+1})(h_{n+1}-h_n),\label{eq:work}\\
 \Eb^{n+1}&=\Eb^n+\delta W_n-(\Es^{n+1}-\Es^n),\label{eq:update}\\
 T_{\rm b}^{n+1}&=\Eb^{-1}(\Eb^{n+1}).\label{eq:invert}
\end{align}
Here $\Eb^{-1}$ denotes the inverse equation of state; for $c_0=0$, $T_{\rm b}=[4\Eb/(Nc_3)]^{1/4}$. Energy absorbed by the resolved subsystem beyond the magnetic work cools the reservoir, whereas heat released into it raises its temperature. The updated temperature controls the next Langevin step, closing the thermal feedback loop (Fig.~\ref{fig:method}).

Heat capacity controls the temperature response, while damping controls energy exchange. A pulse shorter than the external thermal-relaxation time approximates thermal isolation; reversible evolution additionally requires internal equilibration. The local fluctuation-dissipation prescription assumes that $T_{\rm b}$ changes little during one step. We use it as a finite-reservoir approximation: energy balance alone does not guarantee an exact microcanonical stationary distribution. Numerical consistency of the energy and temperature updates is discussed in Appendix~\ref{sec:validation}.

\subsection{Coupled Langevin evolution}
With $\bm F_i=-\partial\mathcal H_{\rm s}/\partial\bm S_i$ and the precession coefficient absorbed into the time unit, the spin equation is
\begin{equation}
 \dot{\bm S}_i=-\bm S_i\times(\bm F_i+\bm\xi_i)
 -\alpha_{\rm s}\bm S_i\times[\bm S_i\times(\bm F_i+\bm\xi_i)],
 \label{eq:spin}
\end{equation}
with independent, zero-mean Gaussian fields satisfying
\begin{equation}
 \av{\xi_i^\mu(t)\xi_j^\nu(t')}=
 \frac{2\alpha_{\rm s}\Tb(t)}{1+\alpha_{\rm s}^2}
 \delta_{ij}\delta_{\mu\nu}\delta(t-t').
 \label{eq:spinnoise}
\end{equation}
The Stratonovich interpretation is used~\cite{Brown1963,GarciaPalacios1998}. There is no additional overall $1/(1+\alpha_{\rm s}^2)$ prefactor in Eq.~\eqref{eq:spin}; this convention fixes the noise normalization. A stochastic Heun predictor and corrector use the same random increment, followed by unit-length normalization.

Each scalar optical coordinate $x$, with velocity $v$ and mass $m_{\rm ph}$, obeys
\begin{align}
 \dot x&=v,\qquad m_{\rm ph}\dot v=f-\zeta v+\eta_x(t),\\
 \av{\eta_x(t)\eta_x(t')}&=2\zeta\Tb(t)\delta(t-t'),\label{eq:latticenoise}
\end{align}
where $f=-\partial\mathcal H_{\rm s}/\partial x$. The mass is $m_{\rm ph}=m_\lambda=K_\lambda/\omega_\lambda^2$ for $\lambda={\rm B,S}$, and the zero-mean optical noises are independent of each other and of the spin noise. In the normalized coordinate of Eq.~\eqref{eq:optical-normalization}, the friction coefficient is $\nu_\lambda=\zeta_\lambda/K_\lambda$, and the normalized random force is $\bar\eta_\lambda=\eta_x/\sqrt{K_\lambda J}$, with covariance $2\nu_\lambda(T_{\rm b}/J)\delta(t-t')$. Consequently, the damping rate is $\gamma_\lambda=\zeta_\lambda/m_\lambda=\nu_\lambda\omega_\lambda^2$. Keeping $\nu_\lambda$ fixed together with $b,b'$ and $\omega_\lambda$ preserves the stochastic dynamics under a change of coordinate normalization. The integration uses a Verlet-type Langevin scheme~\cite{GronbechJensen2013},
\begin{align}
 x_{n+1}&=x_n+b_{\rm G}\Delta t\,v_n
 +\frac{b_{\rm G}\Delta t^2}{2m_{\rm ph}}f_n
 +\frac{b_{\rm G}\Delta t}{2m_{\rm ph}}R_n,\label{eq:gjfx}\\
 v_{n+1}&=a_{\rm G}v_n+\frac{\Delta t}{2m_{\rm ph}}(a_{\rm G}f_n+f_{n+1})
 +\frac{b_{\rm G}}{m_{\rm ph}}R_n,\label{eq:gjfv}\\
 a_{\rm G}&=\frac{1-\zeta\Delta t/(2m_{\rm ph})}{1+\zeta\Delta t/(2m_{\rm ph})},\qquad
 b_{\rm G}=\frac{1+a_{\rm G}}{2},
\end{align}
with $\av{R_n^2}=2\zeta T_{\rm b}^n\Delta t$. The same noise increment enters position and velocity updates, and the old bath temperature is held fixed throughout the step.

Each step computes the spin predictor and new optical positions, advances the field, and corrects the spins. Optical velocities use forces at the corrected spins and new positions. Equations~\eqref{eq:work}--\eqref{eq:invert} then update the reservoir separately in each realization before ensemble averaging.

\subsection{Thermodynamic interpretation}
In the reversible, internally equilibrated limit, the Maxwell relation gives
\begin{equation}
 \left.\frac{\dd T}{\dd h}\right|_{S_{\rm tot}}
 =-\frac{T}{C_{\rm tot}}
 \left(\frac{\partial M}{\partial T}\right)_h.
 \label{eq:maxwell}
\end{equation}
Here $C_{\rm tot}=T(\partial S_{\rm tot}/\partial T)_h$ is the extensive heat capacity of the spins, explicit optical modes, and reservoir. For positive $C_{\rm tot}$, $(\partial M/\partial T)_h$ determines whether the system warms or cools. A negative-to-positive change in this derivative produces a temperature maximum without necessarily indicating a phase transition. A continuous transition can instead produce a slope anomaly without latent heat. Heat capacity affects both the magnitude of the response and the path through $(h,T)$, so changing it need not simply rescale the temperature axis. Even in the reversible limit, the derivative of an adiabatic magnetization curve differs from the isothermal susceptibility.

The reservoir entropy change is
\begin{equation}
 \Sb(T)-\Sb(T_r)=N\left[c_0\ln\frac{T}{T_r}
                  +\frac{c_3}{3}(T^3-T_r^3)\right].
 \label{eq:entropy}
\end{equation}
Here $T_r$ is a reference temperature. In the reversible limit, $\Delta S_{\rm s}=-\Delta\Sb$, with $S_{\rm s}$ including optical and spin degrees of freedom. When a common instantaneous temperature exists but entropy is produced at rate $\dot S_{\rm irr}$, the corresponding relation becomes
\begin{equation}
 \frac{\dd T}{\dd h}=-\frac{T}{C_{\rm tot}}M_T
 +\frac{T}{C_{\rm tot}}\frac{\dot S_{\rm irr}}{\dot h},
 \qquad M_T=\left(\frac{\partial M}{\partial T}\right)_h.
 \label{eq:irreversible}
\end{equation}
Thus a temperature maximum need not occur at $M_T=0$ in a driven process. If subsystem temperatures differ, even this single-temperature description is insufficient.

\subsection{Thermal content of the classical model}\label{sec:heat-capacity}
At fixed temperature and field, the thermal content of the classical model follows exactly by integrating the Gaussian optical positions and momenta in the canonical partition function. There are $6N$ harmonic oscillators, and their spin-dependent centers only change the effective spin Hamiltonian. At fixed stiffness,
\begin{equation}
 Z_{\rm s}(T,h)=A T^{6N}Z_{\rm eff}(T,h),\qquad
 C_{\rm s}=C_{\rm eff}+6N,
 \label{eq:partition}
\end{equation}
where $A$ is independent of $T,h$, and $C_{\rm eff}$ is the heat capacity of the effective spin Hamiltonian containing exchange, DM, $J_{3a}$, the Zeeman term, and Eq.~\eqref{eq:effective}. The additive $6N$ is the classical optical heat capacity, independent of optical masses. Thus the thermal degrees of freedom must be specified alongside the static effective interactions.

In the reversible limit, conservation of the total entropy takes the form
\begin{align}
 0={}&S_{\rm eff}(T,h)-S_{\rm eff}(T_0,h_0)\nonumber\\
 &+N(6+c_0)\ln(T/T_0)+\frac{Nc_3}{3}(T^3-T_0^3).
 \label{eq:isentropic-reference}
\end{align}
Here $S_{\rm eff}$ is the entropy of the effective spin Hamiltonian with the same $D_0,J_{3a},b,b'$. This relation separates the spin, optical, and reservoir contributions to reversible entropy redistribution. The simulations obtain $T_{\rm b}(t)$ directly from energy exchange through Eqs.~\eqref{eq:work}--\eqref{eq:invert}, without imposing an isentropic path.

The persistence of the optical contribution as $T\to0$ is a classical-equipartition artifact when applied to a quantum material: high-energy optical modes freeze out thermally, although they can still generate effective spin interactions. Magnetic quantum corrections additionally depend on the excitation spectrum. Quantum heat-bath approaches offer a route toward appropriate spin and lattice statistics~\cite{Woo2015}. Independently, the adopted $c_3J^3=10^4$ substantially exceeds the acoustic scale inferred from measured lattice heat capacity~\cite{Kant2009}; it specifies a phenomenological thermal background rather than a measured acoustic coefficient. Quantitative reservoir calibration should account for both acoustic and unresolved optical contributions, excluding the modes retained explicitly.

Our aim is to reproduce the qualitative magnetic and thermal response with this simple model. A smooth reservoir without explicit field dependence already allows the plateau, entry cooling, temperature dome, and exit anomaly to emerge together, without prescribing a temperature trajectory. Material-specific amplitudes require quantum heat storage, independent reservoir calibration, and consistent noise and dissipation; robustness to these changes remains to be tested. All explicit classical degrees of freedom are retained in the present energy accounting.

\section{Simulation and experimental data}\label{sec:simulation}
We simulate a periodic lattice with $N=16L^3=1024$ spins, $3N$ bond coordinates, and $3N$ site components. Table~\ref{tab:parameters} lists the common parameters. We compare two optical frequencies, $\omega_{\rm B}/J=\omega_{\rm S}/J=\omega/J=1$ and 10, each at ten initial temperatures $T_0/J=0.050,0.055,\ldots,0.095$. All other model and reservoir parameters are held fixed.

\begin{table}
\caption{Simulation parameters in units $J=k_{\rm B}=1$. Optical frequencies are reported as $\omega_\lambda/J$; the friction coefficients $\nu_\lambda=\zeta_\lambda/K_\lambda$ use the same spin-dynamical time convention. Spring constants only fix the coordinate normalization. The reservoir coefficients specify a phenomenological thermal background and are not fitted to the measured lattice heat capacity.}
\label{tab:parameters}
\begin{ruledtabular}
\begin{tabular}{ll}
Quantity & Value\\
\hline
$J$, $J_{3a}$, $J_{3b}$, $D_0$ & $1$, $-0.03$, $0$, $-0.10$\\
$b$, $b'$, $\eta$ & $0.20$, $0.12$, $0.60$\\
$\omega_{\rm B}/J=\omega_{\rm S}/J$ & $1$ or $10$\\
$\alpha_{\rm s}$, $\nu_{\rm B}$, $\nu_{\rm S}$ & $0.001$, $0.1$, $0.1$\\
$c_0$, $c_3$ & $0$, $10^4$\\
$L$, $N$, field direction & $4$, $1024$, $[111]$\\
Integration step $\Delta t$ & $0.01$\\
Pulse steps & $2^{25}$\\
Independent trajectories per set & $128$\\
\end{tabular}
\end{ruledtabular}
\end{table}
Initial states are annealed linearly from $T/J=10$ to 0.2 over $10^4$ Monte Carlo steps and then to $T_0$ by a cumulative $10^5$ steps, with a total thermalization length of $2^{17}$ steps. One Monte Carlo step is an $N$-site sweep. Every sixth step uses Metropolis updates of the optical-integrated spin energy, followed by Gaussian coordinate sampling; intervening steps overrelax spins and distortions~\cite{GenSuwa2022}. Optical coordinates are drawn from their conditional Gaussian distribution, rather than placed only at their energy minima. For example,
\begin{equation}
 q_{ij}\mid\{\bm S\}\sim\mathcal N\!\left(
 \frac{g_{\rm B}}{K_{\rm B}}\bm S_i\cdot\bm S_j,
 \frac{T}{K_{\rm B}}\right),
 \label{eq:initial-gaussian}
\end{equation}
with the analogous site distribution centered at $(g_{\rm S}/K_{\rm S})\sum_j\bm e_{ij}(\bm S_i\cdot\bm S_j)$ and component variance $T/K_{\rm S}$. At the start of dynamics, optical velocities are Maxwell distributed and the reservoir temperature is $T_0$. This annealing protocol does not itself exclude bias toward metastable initial states.

We use a phenomenological magnetic-field profile inspired by the general shape of single-turn-coil (STC) pulses at the Institute for Solid State Physics, University of Tokyo~\cite{Gen2021,Nomura2021}. We represent this profile by a damped sinusoid,
\begin{equation}
 B_p(\tau)=220e^{-0.12\tau}\cos(0.45\tau-\pi/2)\ {\rm T},
 \label{eq:pulse}
\end{equation}
Typical STC fields reach their maximum within 2--3~$\mu$s and remain positive for approximately 6--8~$\mu$s; coil deformation modifies the field-to-current ratio during the pulse~\cite{Gen2021}. Measurements of the field and its derivative document the rapid, nonuniform sweep~\cite{Nomura2021}. Applications include probing ZnCr$_2$O$_4$ up to 190~T~\cite{Miyata2011} and measuring MCE up to 120~T~\cite{Ogawa2026}. These measurements motivate the shape of Eq.~\eqref{eq:pulse}; the chosen profile is phenomenological and is not fitted to an experimental waveform or sweep rate.

We use pulse-clock increment $\Delta\tau=6.981/2^{25}$ and field normalization $h/J=(2\mu_{\rm B}/k_{\rm B})B_p/(22\ {\rm K})$. The pulse clock $\tau$ is not calibrated in microseconds, and $B_p$ is not a fitted conversion between the simulation and experimental axes. The pulse clock and dynamical time advance with different increments. In reduced dynamical time, $\tau=(\Delta\tau/\Delta t)t$; changing the integration step at fixed physical pulse requires keeping this ratio and total evolution time fixed.

For a physical unit-spin mapping, $J=J_{\rm ex}S^2$, $h=g\mu_{\rm B}SB$, $T_{\rm phys}=(J/k_{\rm B})(T/J)$, and $t_{\rm phys}=[\hbar/(J_{\rm ex}S)]t$. The waveform normalization corresponds to $J_{\rm ex}S/k_{\rm B}=22$~K for $g=2$, giving approximately 16~T per unit $h/J$. This conversion does not align the simulated and experimental transition fields. The exchange estimates below similarly give a time to the pulse maximum of only about 50--60~ns, much shorter than either experimental protocol. In particular, the simulated pulse duration is matched to neither the approximately 8-$\mu$s magnetization pulse nor the 36- and 11-ms MCE pulses used in the comparison. We therefore compare dimensionless feature sequences on separate axes and do not infer experimental relaxation rates from this pulse.

\subsection{Optical-frequency scale}\label{sec:optical-scale}
Infrared spectroscopy finds spin-dependent shifts near 150 and 365~cm$^{-1}$ in CdCr$_2$O$_4$, corresponding to 18.6 and 45.3~meV, with the strongest splitting in the higher branch~\cite{Aguilar2008,Kant2009}. Estimates of the microscopic exchange for Cr spins $S=3/2$ are $J_{\rm ex}=1.02$~meV~\cite{Chung2005} and $J_{\rm ex}/k_{\rm B}=14.7$~K, or 1.27~meV~\cite{Kant2009}. With the unit-spin coefficient $J=J_{\rm ex}S^2$, the precession scale is $J/(\hbar S)$. The optical-to-spin frequency ratio is therefore
\begin{equation}
 r_{\rm ph}=\frac{\Omega_{\rm ph}}{J/(\hbar S)}
 =\frac{\hbar\Omega_{\rm ph}}{J_{\rm ex}S}.
 \label{eq:optical-scale}
\end{equation}
In the spin-dynamical convention, $r_{\rm ph}=\omega/J$. The lower branch gives $r_{\rm ph}=9.8$--12.2, motivating the reduced frequency $\omega/J=10$; the higher branch gives 24--30. This estimate sets a representative dynamical scale without assigning the effective coordinates to particular phonon eigenvectors. Their classical thermal treatment is discussed in Sec.~\ref{sec:heat-capacity}, and Appendix~\ref{sec:frequency} compares $\omega/J=1$ and 10.

\subsection{Experimental comparison}
Experimental data are taken from the published study~\cite{Gen2026}. Magnetization is normalized by its mean over 100--110~T, and its field derivative uses the same factor. Magnetization was measured on a polycrystal initially at 4.2~K with an approximately $8\,\mu$s pulse. MCE measurements used a single crystal with field along $[111]$ and approximately 36~ms and 11~ms pulses in the 60~T and 66~T setups, respectively. The simulated reservoir temperature $T_{\rm b}$ serves as the model counterpart of a thermalized lattice background sensed by thermometry. We compare the sequence of its warming, cooling, and slope changes with experiment, without identifying its magnitude with the measured temperature. Given the different measurement protocols and the phenomenological thermal model, theory and experiment are displayed side by side on their respective field and temperature axes.

Magnetostriction is obtained from the mean bond-phonon displacement. We define the dimensionless bond expansion
\begin{align}
 Q(h)&=\frac{g_{\rm B}}{3NJ}\sum_{\langle ij\rangle}\av{q_{ij}(h)},\nonumber\\
 \left(\frac{\Delta L}{L}\right)_{\rm calc}
 &=A[Q(h)-Q_{\rm ref}],
 \label{eq:magnetostriction}
\end{align}
where $3N$ is the number of nearest-neighbor bonds and $Q_{\rm ref}$ is the low-field reference value. The average linear bond-length change from site displacements cancels in the ideal periodic lattice, although those displacements still affect magnetic ordering and dynamics. The bond average thus provides a scalar measure of sample expansion. The present model does not determine the absolute magnetostriction scale. We therefore choose a single positive, field-independent factor $A\simeq4.7\times10^{-3}$ to match the mean strain within the experimental plateau. The experimental longitudinal strain curves, measured on a polycrystal initially at 4.2~K with 36-ms and $8\,\mu$s pulses, are reproduced from the published vector graphics on their dimensionless $\Delta L/L$ scale~\cite{Gen2026}. This amplitude calibration permits comparison of the field dependence, rather than an independent prediction of the absolute strain.

Pulsed-field simulation curves are trajectory averages, with $\dd m/\dd h$ obtained by numerical differentiation of the mean magnetization. Statistical uncertainty in magnetization, its derivative, and temperature is estimated by bootstrap resampling of independent trajectories; shaded bands denote 95\% confidence intervals. Temperature curves are smoothed for display.

For the isothermal reference in Fig.~\ref{fig:comparison}(a), Monte Carlo simulations use the same interactions, $L=4$, and $[111]$ field direction at $T/J=0.01$. Each sweep direction has 100 annealed runs, with $|\Delta h|/J=0.03$ and $2^{15}$ sweeps each for equilibration and sampling at every field. At each field we select the lowest mean spin energy among these runs in both directions. The resulting curve provides a low-temperature estimate of the expected first-order plateau-entry field, rather than a free-energy determination of coexistence. Its selection uncertainty is not represented by the pulse-trajectory confidence bands.

\section{Results}\label{sec:results}
\subsection{Magnetization, magnetostriction, and magnetocaloric response}
\begin{figure*}
 \paperfigure{0.94\textwidth}{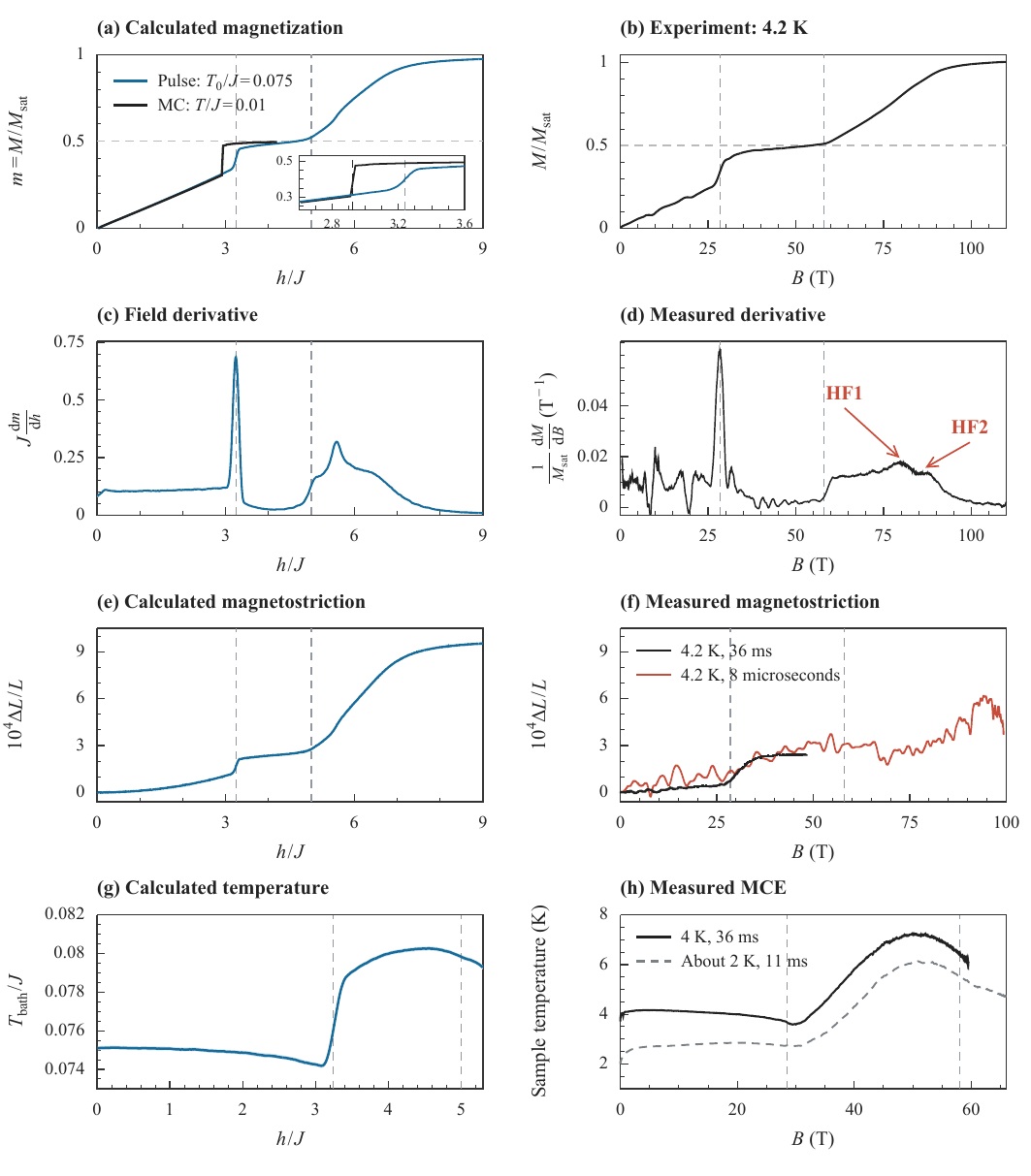}
 \caption{Magnetization, field derivative, magnetostriction, and temperature: simulation at $T_0/J=0.075$, $\omega/J=10$ (blue, left) and experiment~\cite{Gen2026} (right). Panel (a) includes the isothermal MC reference at $T/J=0.01$ (black); its earlier jump is resolved in the inset. Shading in (a), (c), and (g) denotes bootstrap confidence intervals. Gray dashed guides mark plateau entry and approximate exit; field axes are not quantitatively matched. Panels (e),(f) show $\Delta L/L$: the model amplitude is matched to the black experimental plateau, while the published experimental strain scale is retained. In (f), black and red curves extend to approximately 48 and 100~T, respectively (polycrystal, 4.2~K, 36~ms and $8\,\mu$s). Magnetization uses a polycrystal at 4.2~K with an $8\,\mu$s pulse; HF1 and HF2 label its higher-field anomalies. Panel (g) focuses on $h/J\le5.3$; Appendix~\ref{sec:frequency} shows the higher-field temperature continuation. In (h), solid and dashed traces start at 4~K (36~ms) and approximately 2~K (11~ms), respectively, for a single crystal along $[111]$.}
 \label{fig:comparison}
\end{figure*}

Figure~\ref{fig:comparison} compares the representative $T_0/J=0.075$, $\omega/J=10$ simulation with experiment. Both magnetization curves develop a half-magnetization plateau before approaching saturation. The associated temperature responses share a characteristic sequence: cooling near plateau entry, warming to a broad maximum within the plateau, and a change in cooling slope near plateau exit. The simulated temperature changes are smaller, and the plateau occupies a smaller fraction of the field range to saturation than the experimental plateau at approximately 28--58~T. The agreement concerns the sequence of magnetic and thermal features and their locations relative to the plateau.

The central result is that a phenomenological thermal reservoir coupled to the spin-lattice system reproduces the magnetic plateau and the associated sequence of cooling, warming, and cooling together. Its heat capacity is smooth and has no explicit field dependence: neither the temperature dome nor the transition anomaly is prescribed as input. Their emergence through the coupled dynamics is a nontrivial qualitative correspondence with experiment, beyond reproducing magnetization alone.

The low-temperature MC reference in Fig.~\ref{fig:comparison}(a) enters the plateau at a lower field than the driven curve. This comparison motivates the delayed-nucleation scenario discussed in Sec.~\ref{sec:implications}. The reference provides an approximate guide to first-order plateau entry; the differing temperatures and sampling protocols prevent assigning the entire offset to nucleation.

The accompanying lattice response provides a further comparison. Figures~\ref{fig:comparison}(e) and (f) show positive magnetostriction, with an increase near plateau entry followed by a region of weaker field dependence. Although the high-field experimental trace is noisy, it shows renewed expansion above the plateau, in qualitative agreement with the calculated trend. Matching the black-curve plateau amplitude fixes only the displacement scale; the changes in slope and their connection to the magnetic plateau follow from the coupled spin-lattice dynamics. Thus the same trajectory accounts qualitatively for magnetic, structural, and thermal signatures.

\subsection{Magnetization crossing and temperature dependence}\label{sec:crossing}
\begin{figure}[!tbp]
 \paperfigure{\columnwidth}{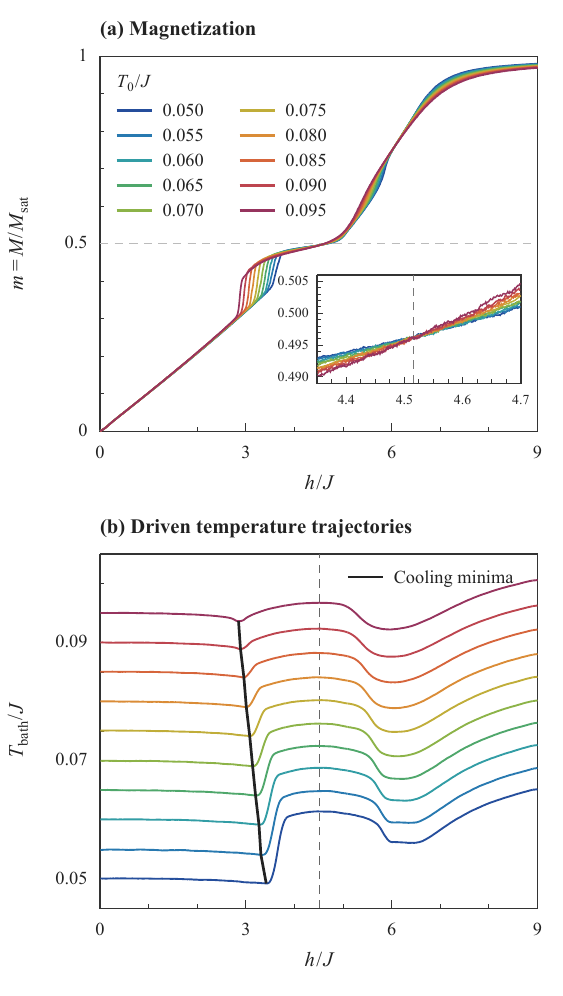}
 \caption{Initial-temperature dependence at $\omega/J=10$. (a) Magnetization with a crossing-region inset; the legend gives $T_0/J=0.050$--0.095 in steps of 0.005. (b) Corresponding reservoir temperatures. The dashed line marks the approximate crossing field $h_\times$. The solid line joins the cooling minima near plateau entry; it is a descriptive guide, not an equilibrium coexistence boundary. Colors agree between panels.}
 \label{fig:crossing}
\end{figure}

Figure~\ref{fig:crossing}(a) compares ten initial temperatures at $\omega/J=10$. The magnetization curves nearly intersect within the plateau, analogous to the reported equilibrium crossing~\cite{Gen2026}. Below the crossing field $h_\times$, trajectories with higher initial temperatures have smaller magnetization; above it, this ordering reverses. The temperature maxima lie near the crossing region. As $T_0$ increases, their height above the initial temperature decreases while their field position changes little. The reservoir parameters remain fixed throughout this temperature series.

Equation~\eqref{eq:maxwell} provides a possible equilibrium interpretation~\cite{Gen2026}: a negative-to-positive change in $(\partial M/\partial T)_h$ changes reversible adiabatic warming into cooling. Our curves are driven trajectories labeled by $T_0$, rather than isotherms. If internal equilibrium holds, the chain rule gives
\begin{equation}
 \left(\frac{\partial m}{\partial T_0}\right)_h
 \simeq\left(\frac{\partial m_{\rm eq}}{\partial T}\right)_h
       \left(\frac{\partial T_{\rm b}}{\partial T_0}\right)_h.
 \label{eq:initial-temperature-chain}
\end{equation}
In this plateau region, $(\partial T_{\rm b}/\partial T_0)_h>0$, so the derivatives would have the same sign under this assumption. The driven curves display the corresponding sign reversal, consistent with the Maxwell interpretation. This correspondence does not independently establish reversibility or bound the entropy-production term in Eq.~\eqref{eq:irreversible}.

The solid line in Fig.~\ref{fig:crossing}(b) connects cooling minima in the instantaneous $(h,T_{\rm b})$ plane. With increasing initial temperature, the minima shift toward lower fields, tracing the thermal signature of plateau entry. This locus need not coincide with either the derivative peaks in Fig.~\ref{fig:comparison} or an equilibrium coexistence boundary, because the sweep can retain metastable states.

\subsection{Anomaly beyond the temperature maximum}\label{sec:plateau-exit}
The experimental MCE anomaly near 58--60~T follows the dome and is associated with the continuous transition from the collinear three-up-one-down plateau to the canted three-plus-one state~\cite{Gen2026}. The $T_0/J=0.075$ simulation also shows a post-peak change in cooling slope as magnetization leaves the plateau, as seen in Fig.~\ref{fig:comparison}(g). The correspondence concerns the sequence relative to plateau exit, without a quantitative mapping of the field axes. Figure~\ref{fig:comparison}(g) focuses on this region; the subsequent cooling at higher fields is retained in Fig.~\ref{fig:frequency}(b) in Appendix~\ref{sec:frequency}. Both optical-frequency families display this structure.

At a continuous transition, changes in heat capacity and the temperature derivative of magnetization can modify the reversible adiabatic slope without latent heat or a magnetization jump. This offers a thermodynamic interpretation of the simulated feature. Our comparison identifies the corresponding change in the magnetic and thermal responses, without assigning a transition order from the temperature trajectory alone.

\subsection{Higher-field magnetic response}\label{sec:hf-phase}
The experimental HF1 and HF2 anomalies around 80--88~T were assigned to an additional phase between the canted state and full polarization~\cite{Gen2026}. In the corresponding equilibrium model, triangular and kagome layers along $[111]$ repeat as T1--K1--T2--K2. The T1 layers have $120^\circ$ transverse order, negative longitudinal spin components, and alternating vector chirality; the other layers are all-up. The kagome layers exhibit breathing distortions, and adjacent layers move toward T1~\cite{Gen2026}. Our simulation shows a broad higher-field response in $\dd m/\dd h$, but does not resolve both boundaries or identify the microscopic order reported in that study.

\section{Relaxation, nucleation, and implications}\label{sec:implications}
Exchange striction changes isotropic exchange strengths but does not break global spin-rotation symmetry. Pairwise conservative exchange torques cancel in the total spin equation, and a fixed-direction Zeeman field has no longitudinal torque. Therefore
\begin{equation}
 \frac{\dd}{\dd t}\sum_i\bm n\cdot\bm S_i=0
 \quad(D_0=0,\ \alpha_{\rm s}=0).
 \label{eq:conservation}
\end{equation}
Optical energy relaxation alone cannot remove this constraint. DM provides a conservative torque that permits longitudinal reorientation, while spin damping and noise also exchange angular momentum with the reservoir. Thus $M$ is not strictly conserved at $\alpha_{\rm s}=10^{-3}$ even without DM. Exploratory simulations motivated the chosen DM interaction and small ferromagnetic $J_{3a}$ because they facilitate plateau entry at weak damping. Both interactions affect the energy landscape as well as the dynamics, so this observation does not identify a unique relaxation mechanism in the material. The present model nevertheless retains the characteristic magnetization plateau and thermal response associated with exchange striction.

Delayed nucleation is one possible source of a narrower plateau. The schematic droplet barrier $\Delta F_*\propto\sigma^3/\Delta f^2$ depends on interfacial energy $\sigma$ and bulk free-energy difference $\Delta f$~\cite{Binder1987}; switching therefore depends on volume, sweep rate, and temperature~\cite{Rikvold1994}. Competing-order barriers can cause slow relaxation even without magnetization conservation~\cite{Oike2025}.

The present lattice is periodic and contains neither defects nor quenched randomness. It therefore lacks surfaces and defect sites that could seed the plateau phase, leaving nucleation to fluctuations within the homogeneous system. A delayed first-order transition can then retain the low-field phase beyond coexistence and shift the magnetization jump upward during the pulse. The earlier jump of the low-temperature MC reference in Fig.~\ref{fig:comparison}(a) is consistent with this interpretation, although it does not independently identify the dynamical nucleation barrier.

Weak defects or quenched disorder that promotes heterogeneous nucleation could reduce this delay. If nucleation enhancement dominates interface pinning and shifts of the phase boundary remain small, the magnetization jump should move closer to coexistence. Earlier entry with little change in plateau exit would widen the plateau toward the experimental shape. This scenario provides a concrete explanation to test for the narrower plateau of the clean periodic simulation, without assuming that disorder makes switching occur exactly at coexistence.

The same mechanism suggests a related change in the thermal response. If earlier plateau formation spreads entropy redistribution over a broader field interval without substantially changing the temperature maximum, the warming slope may become more gradual and closer to the experimental dome. This is a prediction for defect-containing simulations: disorder can also alter heat capacity, latent heat, and irreversible heating, so a shift of the magnetization jump alone does not fix $T_{\rm b}(h)$. Seeded domains, controlled disorder, and variations of system size and pulse rate offer direct tests of the proposed link between plateau width and temperature slope.

The separation of ordering distortions and thermal background may also be useful for kagome magnets~\cite{GenSuwa2022} and MnCr$_2$S$_4$~\cite{Miyata2020,Yamamoto2021,Duc2024}, with phonon calculations guiding the resolved modes~\cite{Kumar2012}. Simultaneous thermometry, magnetization, and structural measurements at several pulse rates would test delayed switching and distinguish orders. Quantitative entropy reconstruction additionally requires calibrated heat capacity, internal equilibration, and irreversible-heating estimates.

\section{Conclusion}\label{sec:conclusion}
We have developed an adiabatic spin-lattice simulation in which explicit bond and site distortions exchange energy with a finite phenomenological reservoir. Updating its temperature from magnetic work and energy exchange produces the experimental sequence of a half-magnetization plateau, entry cooling, a temperature dome, and a change in cooling slope near plateau exit. The joint appearance of these signatures without an imposed temperature profile is the central qualitative result. The near crossing of magnetization curves is consistent with the thermodynamic interpretation of the dome, while the low-temperature MC reference motivates a nucleation-based interpretation of delayed plateau entry.

These conclusions concern the magnetic and thermal trajectories generated by the model. Material-specific temperature amplitudes and entropy estimates remain future work, requiring quantum magnetic and lattice thermodynamics and a calibrated reservoir with compatible noise and dissipation. A natural extension is to include unresolved optical modes alongside the acoustic background, using $c_0>0$ where a constant heat capacity is justified, or a temperature-dependent quantum contribution otherwise. Testing this extension would establish whether the qualitative response persists with a smaller cubic coefficient; it does not by itself resolve the low-temperature quantum heat-capacity mismatch. The classical thermal description sets the scope of the present qualitative comparison.

Numerical refinements include testing time-step convergence and comparing the Heun and GJF schemes with alternative integrators under the same continuous-time pulse. Controlled disorder and pulse-rate studies can address the proposed nucleation mechanism and its thermal signature. More broadly, separating the spin-coupled distortions from the internal thermal background offers a framework for investigating magnetocaloric dynamics in other frustrated spin-lattice systems.

\section*{Acknowledgments}
The author thanks Masaki Gen for fruitful discussions on spin-lattice coupling and the magnetocaloric effect.
Calculations were performed using computational resources of the Supercomputer Center at the Institute for Solid State Physics, the University of Tokyo.
This work was supported by the JSPS KAKENHI Grants-In-Aid for Scientific Research (No. 24H01609).

\appendix
\section{Optical-frequency comparison}\label{sec:frequency}
\begin{figure}[!tbp]
 \paperfigure{\columnwidth}{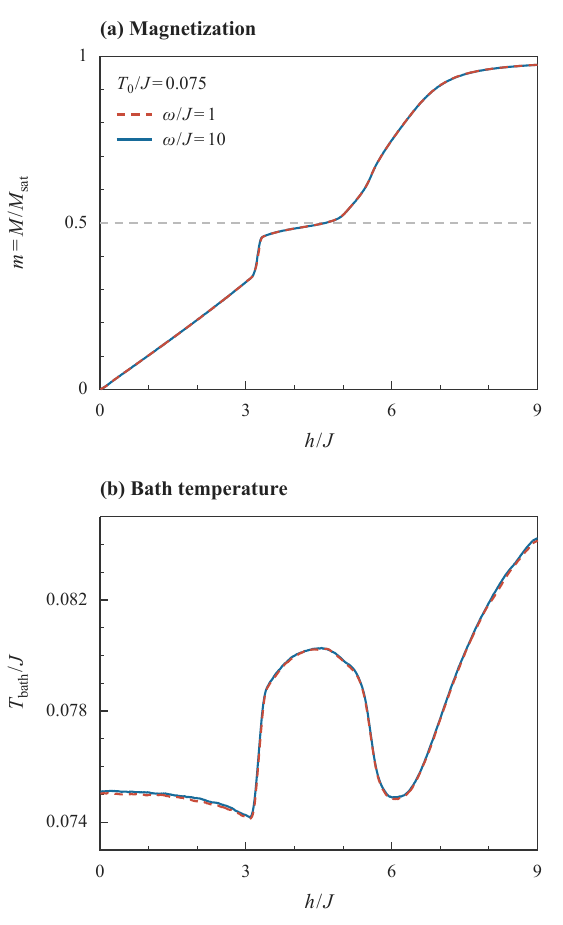}
 \caption{Frequency comparison at $T_0/J=0.075$: (a) magnetization and (b) reservoir temperature for $\omega/J=1$ (red dashed) and 10 (blue solid). The curves use identical normalization, with no relative field shift.}
 \label{fig:frequency}
\end{figure}

Figure~\ref{fig:frequency} compares $\omega/J=1$ and 10 at $T_0/J=0.075$ without shifts or rescaling. The magnetization and temperature curves nearly coincide, including the dome and post-dome structure. These results indicate similar qualitative behavior for the two parameter sets; small differences are not interpreted as resolved physical frequency effects or used to select a better material fit.

The qualitative response is similar, and spectroscopy motivates the representative $\omega/J=10$ dynamical scale. This comparison does not test the classical optical heat capacity, which is independent of frequency. At fixed $\nu_\lambda=\zeta_\lambda/K_\lambda$, increasing $\omega_\lambda$ changes both the damping rate $\gamma_\lambda=\nu_\lambda\omega_\lambda^2$ and the damping ratio $\gamma_\lambda/(2\omega_\lambda)$, as well as $\omega_\lambda\Delta t$. Frequency, damping, and numerical error are therefore not independently varied.

\section{Numerical checks}\label{sec:validation}
For the simulation parameters used here, the audited trajectories satisfy the magnetic-work balance closely, and all saved reservoir temperatures are positive. The checks below quantify this consistency and the known finite-step error of the optical integrator.

The energy-balance residual per spin is
\begin{equation}
 \begin{aligned}
 r_k={}&\frac{\Es(k)+\Eb(k)-\Es(0)-\Eb(0)}{N}\\
 &+\sum_{j<k}\frac{m_j+m_{j+1}}2(h_{j+1}-h_j).
 \end{aligned}
 \label{eq:residual}
\end{equation}
Using samples separated by 128 integration steps, maximum absolute residuals over all 128 trajectories at $T_0/J=0.075$ and 0.08, both with $\omega/J=10$, are of order $10^{-5}J$ per spin. At least one trajectory of every other set was also checked. The saved temperatures of all 2560 trajectories considered here are positive. These results show good energy accounting at the saved times; clipping between outputs cannot be excluded because every-step diagnostics were not recorded.

The GJF on-site velocity has a known fixed-temperature harmonic bias~\cite{GronbechJensen2020},
\begin{equation}
 \left\langle\frac{m_{\rm ph}v_n^2}{2}\right\rangle
 =\frac{T}{2}\left[1-\frac{(\omega\Delta t)^2}{4}\right].
 \label{eq:gjf-bias}
\end{equation}
Independent discrete-covariance calculations reproduce this expression. At the adopted $\Delta t=0.01$, the kinetic-energy deficits are only 0.0025\% for $\omega/J=1$ and 0.25\% for $\omega/J=10$. An independent undamped harmonic-oscillator test at fixed duration also shows systematic error reduction: halving the time step reduces the maximum relative energy deviation by approximately a factor of four. For $\omega/J=10$, these deviations are $2.5\times10^{-3}$, $6.25\times10^{-4}$, and $1.56\times10^{-4}$ at steps 0.01, 0.005, and 0.0025, respectively.

The harmonic benchmarks establish small finite-step errors in these limiting cases, rather than an error bound for the coupled temperature trajectory. In particular, Eq.~\eqref{eq:update} transfers a resolved-energy error $\epsilon$ into the reservoir, giving $\delta T_{\rm b}\simeq-\epsilon/C_{\rm b}$ even when the total work balance closes. Energy accounting and temperature convergence are therefore complementary checks.

The available diagnostics support the energy bookkeeping and quantify a small known harmonic-integrator bias at the chosen parameters. The central physical result is the joint appearance of the magnetic plateau, entry cooling, temperature dome, and exit anomaly in the computed trajectories; it does not rely on resolving tiny frequency-dependent differences or precise temperature amplitudes. These qualitative observations are the focus of the present study. Establishing that they are unchanged by discretization still requires a time-step convergence test of the coupled spin-lattice system with the same continuous-time pulse.

\bibliography{references}
\end{document}